\documentclass[prl,twocolumn]{revtex4-1}
\usepackage{amsmath,graphicx}
\usepackage{hyperref}
\hypersetup{colorlinks=true, urlcolor=blue,
  linkcolor=blue, citecolor=blue}

\renewcommand{\figurename}{\textbf{Fig.}}

\begin{document}

\title{Non-Abelian Braiding of Chiral Majorana Fermions by Quantum Dots}

\author{Yan-Feng Zhou$^{1,2}$}
\author{Zhe Hou$^{1,2}$}
\author{Qing-Feng Sun$^{1,2,3}$}\email{sunqf@pku.edu.cn}
\affiliation{$^{1}$International Center for Quantum Materials, School of Physics, Peking University, Beijing 100871, China}
\affiliation{$^{2}$Collaborative Innovation Center of Quantum Matter, Beijing 100871, China}
\affiliation{$^{3}$CAS Center for Excellence in Topological Quantum Computation, University of Chinese Academy of Sciences, Beijing 100190, China}
\date{\today}
\begin{abstract}
The non-Abelian braiding of Majorana fermions is one of the most promising operations
providing a key building block for the realization of topological quantum computation.
Recently, the chiral Majorana fermions were observed in a hybrid junction btween
a quantum anomalous Hall insulator and an s-wave superconductor.
Here we show that if a quantum dot or Majorana zero mode couples to the chiral Majorana fermions,
the resulting resonant exchange of chiral Majorana fermions can lead to the non-Abelian braiding.
Remarkably, any operation in the braid group can be achieved by this scheme.
We further propose electrical transport experiments to observe the braiding of
four chiral Majorana fermions and demonstrate the non-Abelian braiding
statistics in four-terminal devices of the hybrid junctions.
Both a conductance peak due to the braiding and
the braiding-order dependent conductance are predicted.
These findings pave a way to perform any braiding operation of chiral Majorana fermions
by electrically controllable quantum dots.
\end{abstract}

\maketitle

Majorana fermion with antiparticle being itself,
is originally introduced as a putative elementary particle by Ettore Majorana,
and has been now pursued as quasiparticle excitation in condensed-matter systems \cite{Alicea2,Beenakker,Sarma2}.
Quantum information can be stored nonlocally in the degenerate ground state space
generated by the zero-energy Majorana excitations.
Because of the non-Abelian braiding statistics \cite{ReadN,Ivanov},
that information can be manipulated through the exchange of the Majorana excitations
which leads to a noncommutative transformation between different ground states.
The final state is determined by the topology of the braiding and is
robust against local perturbation,
with possible applications in topological quantum computation \cite{AKitaev,Freedman,Sarma1,CNayak,Alicea1}.

For the zero-dimensional case, Majorana zero modes (MZMs) are predicted
as midgap states bound to charge-$e/4$ quasiparticles of $\nu=5/2$ fractional
quantum Hall effect \cite{ReadN,GMoore,Radu,Dolev}
and Abrikosov vortices in topological superconductors (TSCs) \cite{Volovik,Ivanov,FuL1,Jia1}.
Moreover, alternative proposals suggest that a semiconducting nanowire coupled
with a superconductor can also support MZMs localized at the wire ends \cite{Kitaev1,Lutchyn,Oreg}
and mounting experimental progress in pursuing MZMs in these systems
has been achieved by measuring the zero-bias peak of tunneling spectroscopy \cite{Mourik,DasA,DengMT,ZhangH}.
Despite the architectures proposed for performing the braiding operations of MZMs \cite{Alicea1,Hyart,Plugge,Karzig},
the experimental realization remains an ongoing challenge.

\begin{figure}
\centering
\includegraphics[width=0.46\textwidth]{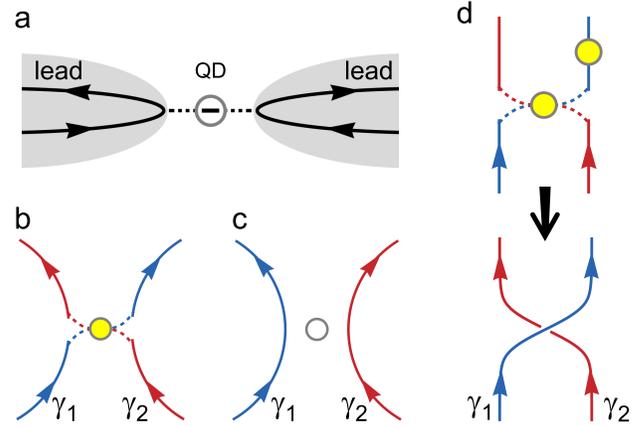}
\caption{\textbf{Braiding of chiral Majorana fermions coupled to MZMs.}
With inspiration of the resonant tunneling in the lead/QD/lead model depicted in \textbf{a},
a system consisting of two chiral Majorana fermions coupled to MZMs (yellow solid circle) is proposed
to braid chiral Majorana fermions as shown in \textbf{b} and \textbf{d}.
When the MZM is absent or disabled as indicated by a hollow circle in \textbf{c},
the chiral Majorana fermions $\gamma_1$ (blue arrow) and $\gamma_2$ (red arrow)
propagate independently. With the presence of a MZM,
a resonant exchange sends $\gamma_1\rightarrow-\gamma_2$ and $\gamma_2\rightarrow-\gamma_1$ as
shown in \textbf{b}. Moreover, a single chiral Majorana fermions coupled
to a MZM will acquire a negative sign.
The braiding according to $\gamma_1\rightarrow\gamma_2$
and $\gamma_2\rightarrow-\gamma_1$ can be realized if one couples two MZMs to $\gamma_{1,2}$ as shown in \textbf{d}.
}
\end{figure}

As one-dimensional analogue of MZMs,
chiral Majarona fermions emerge as unidirectionally propagating edge modes
surrounding the edge of $p+ip$ TSC \cite{QiXL1,Mackenzie,FuL1,SauJD,WangJ,YZhou}.
The TSC has a full pairing gap classified by topological Chern number $\mathcal{N}$
which also determines the number of chiral Majorana edge modes.
Theoretical proposals show that such an exotic superconductivity could arise
by bringing certain topological matters in proximity to an s-wave superconductor \cite{QiXL1,FuL1}.
By coupling $\mathrm{Nb}$ superconductor to a quantum anomalous Hall insulator (QAHI)
realized in magnetic topological insulator thin films \cite{ChangCZ,Kandala},
a half-integer conductance plateau resulting from the chiral Majarona fermion
was experimentally observed \cite{HeQL}, serving as a hopeful signature of TSC with $\mathcal{N}=1$.
Recently, one group \cite{LianB} recognized
that the propagation of chiral Majarona fermions can lead to
braiding for performing topological quantum computations similar to the braiding of MZMs.
In contrast to bounded MZM, the chiral Majorana fermion is extended
providing a promising platform for scalable quantum computations.

Motivated by the resonant tunneling of an electron through a quantum dot (QD)
in the lead/QD/lead system \cite{Beenakker2} as shown in Fig.1\textbf{a},
we first show that if two chiral Majarona fermions $\gamma_{1,2}$ are coupled to a MZM,
the resonant tunneling of Majarona fermions occurs and leads to
that a resonant exchange sends $\gamma_1\rightarrow -\gamma_2$
and $\gamma_2 \rightarrow -\gamma_1$ (see Fig.1\textbf{b}).
On the other hand, a single chiral Majarona fermion coupled with a MZM
acquires a negative sign.
Our key idea is that by combining these two effects due to the coupling with MZMs (see Fig.1\textbf{d}),
$\gamma_1$ and $\gamma_2$ are transformed according to $\gamma_1\rightarrow \gamma_2$
and $\gamma_2 \rightarrow -\gamma_1$, reminiscent of the braiding of MZMs.
Then we demonstrate that such braiding of chiral Majarona fermions
could be completely reproduced by replacing MZMs with QDs,
which are well constructed and can be controlled by gate voltage in experiments\cite{Petta,Koppens,Zwanenburg}.
We further propose a four-terminal setup of the hybrid TSC/QAHI junction to observe
the braiding of four chiral Majarona fermions indicated by a resonant conductance peak.
Remarkably, our proposed system can be extended to carry out
any braiding operation in the braid group.
For sequential exchanging process,
the resonant conductance peak has a value of $\frac{e^2}{2h}$ or $\frac{e^2}{h}$
depending on the exchanging order which means the non-Abelian braiding statistics.

We begin by investigating the MZM coupling case
in which a MZM is coupled to a pair of chiral Majarona fermions as shown in Fig.1\textbf{b}.
The low energy Hamiltonian of two decoupled chiral Majarona fermions is
\begin{equation}\label{CMF}
  H_0=i\nu\sum_{\alpha=1,2}\int_{-\infty}^{+\infty}\gamma_\alpha(x)\partial_x\gamma_\alpha(x)dx,
\end{equation}
where $\gamma_\alpha(x)$ is the field operators of chiral Majarona fermions with
$\gamma_\alpha^\dag(x)=\gamma_\alpha(x)$ satisfying
$\{\gamma_\alpha(x),\gamma_\beta(x')\}=\delta_{\alpha\beta}\delta_{xx'}$,
and $\nu$ denotes the Fermi velocity.
Let $H_\mathrm{M}$ describe the coupling term between the MZM
$\gamma_0$ ($\gamma_0^\dag=\gamma_0$ and $\{\gamma_0,\gamma_0\}=1$)
and $\gamma_\alpha(x)$ at $x=0$ with a strength $t_\alpha$.
Then the total Hamiltonian is
\begin{equation}\label{H1}
  H_1=H_0+H_{\mathrm{M}},
\end{equation}
where $H_{\mathrm{M}}=\sum_{\alpha=1,2}it_\alpha\gamma_\alpha(0)\gamma_0$.

When MZM $\gamma_0$ is absent as denoted by a hollow circle in Fig.1\textbf{c},
$\gamma_1$ and $\gamma_2$ propagate independently determined by $H_0$.
If we switch on $\gamma_0$ as indicated by a yellow solid circle in Fig.1\textbf{b},
the scattering between $\gamma_1$ and $\gamma_2$ occurs at $x=0$
due to the coupling with a MZM.
In order to study the transport process, we first calculate the scattering
matrix of the chiral Majarona fermions described by $H_1$ in Eq.(\ref{H1}).
Denoting the incoming and outgoing scattering states of the chiral Majarona fermions
by $\gamma_{1/2}(0^-)$ and $\gamma_{1/2}(0^+)$, respectively,
the scattering matrix can be written as \cite{suppl}
\begin{equation}\label{SM1}
  \left(
    \begin{array}{c}
      \gamma_1(0^+) \\
      \gamma_2(0^+) \\
    \end{array}
  \right)=S_\mathrm{M}\left(
               \begin{array}{c}
                 \gamma_1(0^-) \\
                 \gamma_2(0^-) \\
               \end{array}
             \right),
\end{equation}
where
\begin{equation}\label{SM2}
  S_\mathrm{M}=\frac{1}{A}\left(
        \begin{array}{cc}
          4i\varepsilon\nu+t_2^2-t_1^2 & -2t_1t_2 \\
          -2t_1t_2 & 4i\varepsilon\nu+t_1^2-t_2^2 \\
        \end{array}
      \right),
\end{equation}
in which $A=4i\varepsilon\nu+t_1^2+t_2^2$ and $\varepsilon$ is the incident energy.
The off-diagonal elements of $S_\mathrm{M}$ matrix correspond to
the amplitude for transmission between $\gamma_1$ and $\gamma_2$.
The resulting transmission coefficient is $T(\varepsilon) =
|S_{\mathrm{M},12}|^2 =\frac{4\Gamma_1\Gamma_2}{4\varepsilon^2+(\Gamma_1+\Gamma_2)^2}$
with $\Gamma_\alpha \equiv t_\alpha^2/(2\nu)$
which is same with the Breit-Winger formula describing the resonant scattering
of the lead/QD/lead system.

Similar to the resonant tunneling process in the lead/QD/lead system \cite{Beenakker2} (see Fig.1\textbf{a}),
if the value of $\varepsilon$ largely deviates from zero,
the energy mismatch disables MZM $\gamma_0$ and no transmission happens
with $T=0$ as illustrated in Fig.1\textbf{c}.
However, it is obvious from Eq.(\ref{SM2})
that when $\varepsilon=0$ and $t_1=t_2=t$,
the scattering matrix becomes $S_\mathrm{M}=\left(
                                                  \begin{smallmatrix}
                                                    0 & -1 \\
                                                    -1 & 0 \\
                                                  \end{smallmatrix}
                                                \right)
$. In other words,
when the energy of the incoming chiral Majarona fermions matches the energy of the MZM,
there occurs a resonant exchange according
to $\gamma_1\rightarrow -\gamma_2$ and $\gamma_2 \rightarrow -\gamma_1$ as depicted in Fig.1\textbf{b}.
Moreover, we consider the case that $\gamma_0$ is only coupled with $\gamma_1$
by setting $t_2=0$ and $t_1=t$.
Here, for $\varepsilon=0$, the scattering
matrix becomes $S_\mathrm{M}=\left(
                                                  \begin{smallmatrix}
                                                    -1 & 0 \\
                                                    0 & 1 \\
                                                  \end{smallmatrix}
                                                \right)
$. This means that the phase of chiral Majarona fermion can be changed by a value of $\pi$
as $\gamma_1\rightarrow-\gamma_1$,
which can be also realized by the coupling with a metallic island \cite{YZhou}.
More interestingly, it can be shown that two successive manipulations
in which one MZM is first coupled simultaneously to $\gamma_1$ and $\gamma_2$,
and then an additional one is coupled solely to $\gamma_1$ (see Fig.1\textbf{d}),
can recover the braiding of chiral Majarona fermions,
$\gamma_1\rightarrow \gamma_2$ and $\gamma_2 \rightarrow -\gamma_1$,
as that in the braiding of MZMs.
At this point, it suggests that our method makes the braiding of chiral Majarona fermions
possible with the coupling of MZMs.

\begin{figure}
\centering
\includegraphics[width=0.5\textwidth]{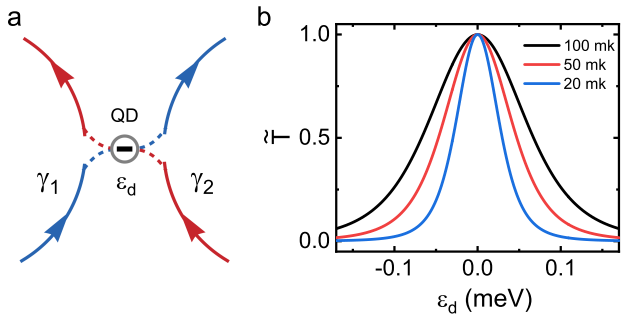}
\caption{\textbf{Electrical control of the braiding of chiral Majorana fermions coupled to a QD.}
Considering that a zero-energy charged fermion is topologically equivalent to a pair of MZMs, a system consisting of
two chiral Majorana fermions coupled
to a QD with a single energy level $\varepsilon_d$ in \textbf{a}
is demonstrated to be able to braid chiral Majorana fermions as well as in Fig.1\textbf{c}.
\textbf{b}, The transmission $\widetilde{T}$ describing the scattering between $\gamma_1$
and $\gamma_2$ due to the QD coupling as a function of $\varepsilon_d$ at different temperatures.
In real experiments, $\varepsilon_d$ inside the QD can be tuned by gate voltages.
For $\varepsilon_d=0$, the resonant exchange of $\gamma_1$ and
$\gamma_2$ occurs ($\widetilde{T}=1$).
}
\end{figure}

Considering that a zero-energy charged fermion bound state is topologically
equivalent to a pair of MZMs \cite{Alicea2},
whether a QD with a single energy level can provide an alternative approach
to braid the chiral Majarona fermions as well as MZM.
Next, we consider that two chiral Majarona fermions are coupled to a QD as shown in Fig.2\textbf{a}.
The total Hamiltonian now becomes
\begin{equation}\label{H2}
  H_2=H_0+H_{\mathrm{QD}}+H_\mathrm{C},
\end{equation}
where $H_{\mathrm{QD}}=\varepsilon_dd^\dag d$ and $H_\mathrm{C}=\sum_{\alpha=1,2}i(\tilde{t}_\alpha/\sqrt{2})\gamma_\alpha(0)(d+d^\dag)$ \cite{LawKT}.
The second term $H_{\mathrm{QD}}$ is the Hamiltonian of the QD
with a single energy level $\varepsilon_d$ and $d^\dag$ ($d$)
are the creation (annihilation) operators of the fermion state in the QD.
The third term $H_\mathrm{C}$ describes the coupling between
$\gamma_{\alpha}$ and the QD with a coupling strength $\tilde{t}_\alpha$.

\begin{figure*}
\centering
\includegraphics[width=0.8\textwidth]{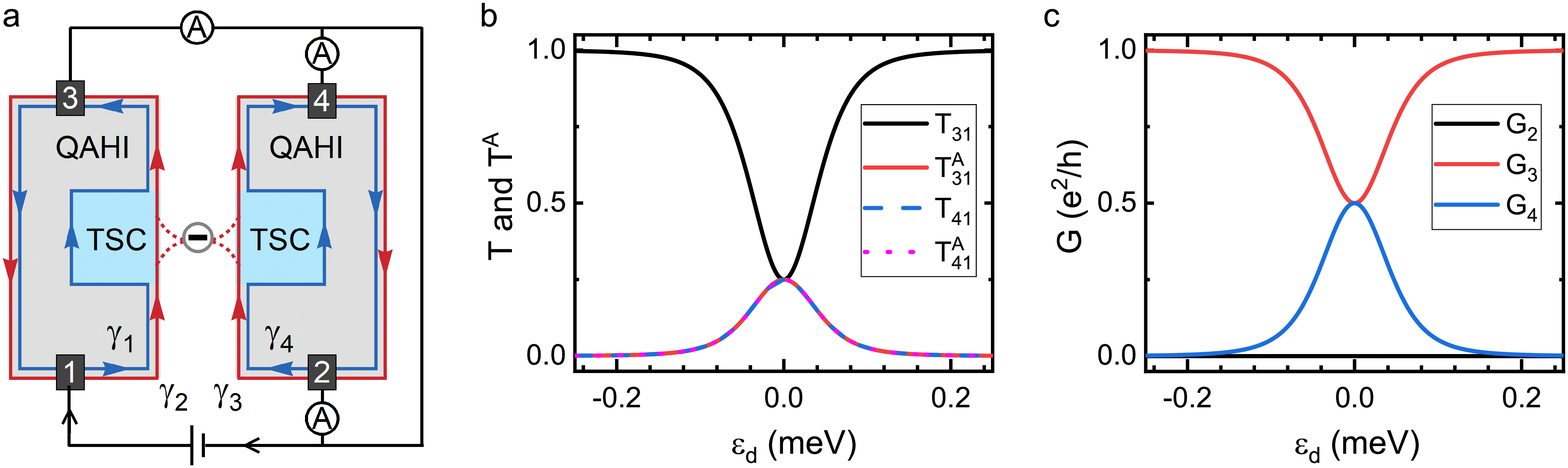}
\caption{
\textbf{Experimental device that allows the observation of the braiding based on TSC/QAHI junctions.}
\textbf{a}, Transport process via four chiral Majorana fermions $\gamma_i$ ($i=1,2,3,4$).
The chirality of $\gamma_i$ in left part is different with
the one in right part resulting from the opposite magnetization in two regions.
The QD coupled to the TSC/QAHI junctions can trigger the exchange of $\gamma_2$ and $\gamma_3$.
\textbf{b} and \textbf{c}, Transport coefficients for the transmission process for lead 1
to lead 3 and 4, and terminal conductance as functions of $\varepsilon_d$.
For $\varepsilon_d=0$, the resonant exchange of $\gamma_2$ and $\gamma_3$ occurs
with $T_{31}=T_{31}^A=T_{41}=T_{41}^A=1/4$, and $G_4$ ($G_3$) shows a peak (valley).
}
\end{figure*}

In this case, the scattering matrix denoted by $S_\mathrm{Q}$ becomes \cite{suppl}
\begin{equation}\label{SQ}
  S_\mathrm{Q}=\frac{1}{\widetilde{A}}\left(
        \begin{array}{cc}
          \widetilde{B}+\tilde{t}_2^2-\tilde{t}_1^2 & -2\tilde{t}_1\tilde{t}_2 \\
          -2\tilde{t}_1\tilde{t}_2 & \widetilde{B}+\tilde{t}_1^2-\tilde{t}_2^2 \\
        \end{array}
      \right),
\end{equation}
where $\widetilde{B}=4i\nu(\varepsilon^2-\varepsilon_d^2)/\varepsilon$ and  $\widetilde{A}=\widetilde{B}+\tilde{t}_1^2+\tilde{t}_2^2$.
For comparison with the MZM setting $\varepsilon_d=0$,
it can be found that the scattering matrix $S_\mathrm{Q}$
is the same as $S_\mathrm{M}$ in Eq.(\ref{SM2}).
Therefore, the QD should be able to accomplish the braiding
of chiral Majarona fermions as the MZM.
On the other hand, in real experiments the energy level $\varepsilon_d$
inside the QD can be tuned easily by gate voltages \cite{Petta,Koppens,Zwanenburg}.
For finite $\varepsilon_d$, the transmission coefficient now becomes
$\widetilde{T}(\varepsilon)=|S_{\mathrm{Q},12}|^2 =\frac{4\widetilde{\Gamma}_1\widetilde{\Gamma}_2}
{4(\varepsilon^2-\varepsilon_d^2)^2+(\widetilde{\Gamma}_1+\widetilde{\Gamma}_2)^2}$
with $\widetilde{\Gamma}_\alpha \equiv \tilde{t}_\alpha^2/(2\nu)$.
Considering the non-zero temperature $\mathcal{T}$,
the effective transmission coefficient is
$\widetilde{T}=\int_{-\infty}^{+\infty}\widetilde{T}(\varepsilon)(-\frac{\partial f}{\partial\varepsilon})d\varepsilon$,
in which $f(\varepsilon)=[\mathrm{exp}(\varepsilon/k_B\mathcal{T})+1]^{-1}$
is the Fermi distribution function.
Fig.2\textbf{b} shows $\widetilde{T}$ as functions of $\varepsilon_d$ with $\widetilde{\Gamma}_1=\widetilde{\Gamma}_2\equiv\Gamma=1\ \mathrm{meV}$
at different temperatures.
It can be seen that the transmission curves $\widetilde{T}$ show
obviously resonant behaviors with a peak at $\varepsilon_d=0$
where the resonant exchange of chiral Majorana fermions occurs.
If the value of $\varepsilon_d$ deviates from zero,
$\widetilde{T}$ decreases with a full-width at half-maximum estimated
to be $\sqrt{\Gamma k_B\mathcal{T}}$.
When $\varepsilon_d$ is far from zero, $\widetilde{T}$ tends to zero
and the chiral Majarona fermions propagate independently without
exchange as shown in Fig.1\textbf{c}.
This result implies an electrical method to control the braiding.
In contrast to MZMs, the QDs are well constructed experimentally \cite{Zwanenburg}
and thus the QDs are considered in the following discussion.

To observe the braiding of chiral Majarona fermions experimentally,
we propose the electrical transport in a four-terminal device
where two ribbons of hybrid TSC/QAHI junctions with opposite out-plane magnetization
are coupled by a QD as depicted in Fig.3\textbf{a}.
The QAHI was realized in magnetically doped topological insulators
with Chern number $\mathcal{C}=1$ and has one chiral Dirac edge mode around its boundary \cite{ChangCZ,Kandala}.
As discussed in the introduction,
the recently observed TSC has a Chern number $\mathcal{N}=1$
which corresponds to one chiral Majorana edge mode \cite{HeQL}.
In view of the fact that a chiral QAHI edge mode is equivalent
to two chiral Majorana edge modes \cite{QiXL1},
the transport process in the device occurs via
four chiral Majorana edge states $\gamma_i$ ($i=1,2,3,4$)
as denoted by red and blue arrows in Fig.3\textbf{a}.
The QD between the TSC regions behaves as a switch for controlling
the braiding of $\gamma_2$ and $\gamma_3$ and determines the terminal conductance.

The measured current in lead $n$ can be calculated using
the multiprobe Landauer-B\"{u}ttiker formula \cite{Sun1,Sun2,datta}
\begin{equation}\label{Cur}
I_n=\frac{e^2}{h}\sum_mT_{nm}(V_n-V_m)+T_{nm}^A(V_n+V_m),
\end{equation}
where $T_{nm}$ ($T_{nm}^A$) is the normal tunneling (Andreev reflection) coefficient
from lead $m$ to lead $n$ and $V_n$ is the voltage of terminal $n$.
The voltage of lead 1 is fixed to $V$ and the voltages
of lead 2,3,4 have the same value (see Fig.3\textbf{a}).
These transport coefficients can be calculated from the scattering
matrix $S_\mathrm{Q}$ in Eq.(\ref{SQ})
(refer to Supplementary Information \cite{suppl} for a detailed derivation of these quantities).
The conductance of lead $n$ is defined as $G_n=\frac{-I_n}{V_1-V_n}$.
Fig.3\textbf{b} and 3\textbf{c} display the transport coefficients and
terminal conductance as functions of $\varepsilon_d$, respectively.
If $\varepsilon_d$ is tuned away from zero,
the electrical transports in two ribbons are independent despite the QD ($T_{41}=T^A_{41}=0$)
and the normal tunneling process dominates with $T_{31}=1$,
leading to $G_3=e^2/h$ and $G_4=0$.
As $\varepsilon_d$ approaches to zero,
the transmission between $\gamma_2$ and $\gamma_3$ takes place with
a resonant exchange $\gamma_2\rightarrow -\gamma_3$ and $\gamma_3 \rightarrow -\gamma_2$.
In this situation, originating from lead 1, $\gamma_1$ enters into lead 3 normally
and $\gamma_2$ tunnels into lead 4 due to the resonant exchange.
As a result, the normal tunneling coefficient and Andreev reflection coefficient
from lead 1 to both lead 3 and 4 are equal, i.e., $T_{31}=T_{31}^A=T_{41}=T_{41}^A=1/4$ as shown in
Fig.3\textbf{b}.
Moreover, it can be seen from Fig.3\textbf{c} that
the conductance $G_4$ shows a peak of $\frac{e^2}{2h}$
while $G_3$ has a valley of $\frac{e^2}{2h}$
with $G_2+G_3+G_4=\frac{e^2}{h}$ and $G_2=0$.

In principle, there are infinite braiding operations
on four chiral Majorana fermions forming a braid group.
Any braiding operation in the group can be represented algebraically
in terms of generators $\sigma_{i}$ with $i=1,2,3$ \cite{CNayak}.
The braiding of $\gamma_\alpha$ at the $i$th position and $\gamma_\beta$ at
($i+1$)th position is a non-Abelian unitary transformation with the form $\sigma_i=\mathrm{exp}(\pi\gamma_\beta\gamma_\alpha/2)$.
Now, we take the device with two ribbons of TSC/QAHI junctions in Fig.4\textbf{a}
as an example to show our strategy to execute all the three generators.
There are four chiral Majorana edge modes denoted by arrowed lines starting from leads 1 and 2.
Similar to the discussion in Fig.1\textbf{d}, with the coupling of two successive QDs,
the consequent operation completes $\sigma_2$
which braids the chiral Majorana fermions propagating along the central two red lines.
Moreover, gate voltage $V_G$ of the left ribbon in Fig.4\textbf{a} can induces
an additional phase for the chiral QAHI edge state leading to
a transformation between the chiral Majorana fermions on the first and second lines
sorted from left to right, which is equivalent to the braiding operator $\sigma_1$ \cite{LianB}.
Similarly, generator $\sigma_3$ can be carried out by placing a gate voltage
on the QAHI edge of the right ribbon.
Accordingly, in view of the scalability of chiral Majorana edge modes,
the proposed device provides a scalable platform to
perform any braiding operation by an arbitrary combination of the three generators.
Moreover, all the braiding operations can be well controlled and tuned by the electrical method \cite{suppl}.

With this exciting possibility to carry out any braiding operation,
we next propose the electrical transport experiments in the devices as shown in Fig.4
to observe the non-Abelian braiding statistics
where the sequential exchanges are executed in different orders.
Here, we set $\varepsilon_d=0$ for all the QDs.
Let us define the occupation number 0 or 1 of the QAHI edge states
in the left and right ribbons as two qubits $L$ and $R$ with
bases $|0_x\rangle$ and $|1_x\rangle$ ($x=L,R$) \cite{LianB}.
The degenerate ground-state space of the two qubits is expanded by four states,
$|0_L0_R\rangle$, $|1_L0_R\rangle$, $|0_L1_R\rangle$, and $|1_L1_R\rangle$.
First, we consider two joint operators $\sigma_2\sigma_1$ and $\sigma_1\sigma_2$.
After the braiding operations, the chiral Majorana fermions ($\gamma_1,\gamma_2,\gamma_3,\gamma_4$)
are transformed to ($-\gamma_2,-\gamma_3,\gamma_1,\gamma_4$) by $\sigma_2\sigma_1$
and to ($\gamma_3,\gamma_1,\gamma_2,\gamma_4$) by $\sigma_1\sigma_2$, respectively.
If we prepare the system into an initial state $|\psi_i\rangle=|1_L0_R\rangle$
by injecting the electrons into the qubit $L$ one by one from lead 1 with weak current,
it can be found that $\sigma_2\sigma_1$ turns the system into a final state $|\psi_{f1}\rangle=(|1_L0_R\rangle-i|0_L1_R\rangle)/\sqrt{2}$ and
$\sigma_1\sigma_2$ turns it into another different state $|\psi_{f2}\rangle=(|1_L0_R\rangle-|0_L1_R\rangle)/\sqrt{2}$, correspondingly \cite{Ivanov}.
Unfortunately, the two different final states cannot be distinguished
by the proposed four-terminal device in which we find $G_3=G_4=\frac{e^2}{2h}$
for both cases \cite{suppl}.
Although the two joint operators are carried out in different order,
they both transport $\gamma_1$ and $\gamma_2$ coming from lead 1 to different leads
(lead 3 and lead 4), respectively, and this leads to the same result of conductance measurements.

However, the operators constituted by three sequential exchanges $\sigma_2\sigma_1\sigma_2$
and $\sigma_2\sigma_2\sigma_1$ in Fig.4 yields very different results,
indicating the non-Abelian braiding statistics.
In this situation, ($\gamma_1,\gamma_2,\gamma_3,\gamma_4$) are
transformed to ($\gamma_3,-\gamma_2,\gamma_1,\gamma_4$) by $\sigma_2\sigma_1\sigma_2$
and to ($-\gamma_2,-\gamma_1,-\gamma_3,\gamma_4$) by $\sigma_2\sigma_2\sigma_1$
as shown in the middle part of Fig.4.
From the initial state $|\psi_i\rangle=|1_L0_R\rangle$,
the device in Fig.4\textbf{a} arrives at the final state
$|\psi_{f2}\rangle=(|1_L0_R\rangle-|0_L1_R\rangle)/\sqrt{2}$
transformed by $\sigma_2\sigma_1\sigma_2$
while the joint operator $\sigma_2\sigma_2\sigma_1$ drives the device in Fig.4\textbf{b}
into a final state $|\psi_{f3}\rangle=|0_L1_R\rangle$ \cite{Ivanov}.
As a result, the conductances observed in Fig.4\textbf{a} are $G_3=G_4=\frac{e^2}{2h}$,
but the ones in Fig.4\textbf{b} are $G_4=\frac{e^2}{h}$ and $G_3=0$ \cite{suppl}.
In the case of $\sigma_2\sigma_2\sigma_1$,
$\gamma_1$ and $\gamma_2$ enter into lead 3 together,
and recombine as a hole providing a different conductance measurement.
These results provide a signature supporting the non-Abelian braiding statistics
of chiral Majarona fermions.

\begin{figure}
\centering
\includegraphics[width=0.47\textwidth]{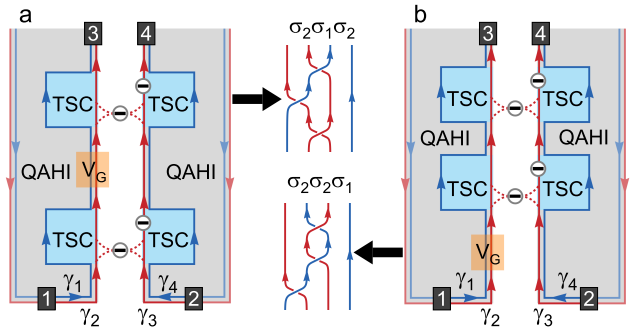}
\caption{
\textbf{Experimental devices providing a demonstration for the non-Abelian braiding statistics.}
To observe the non-Abelian statistics of the braiding in the electrical transport,
one has to perform sequential exchanges and demonstrate that the measurement results of the device
are dependent on the braiding order.
The gate voltage $V_G$ performs the operator $\sigma_1$ and the operator $\sigma_2$
is carried out by two successive QDs as discussed in Fig.1\textbf{d}.
Two joint operations constituted by three sequential exchanges,
$\sigma_2\sigma_1\sigma_2$ as realized in \textbf{a}
and $\sigma_2\sigma_2\sigma_1$ as realized in \textbf{b},
turn the systems into completely different final states
and yield different measurement results of the terminal conductances.
The middle part shows the corresponding braiding of $\gamma_i$ ($i=1,2,3,4$) in the two devices.
}
\end{figure}

To conclude, we proposed a new method to perform
any braiding operation of chiral Majarona fermions coupled with QDs or MZMs
and proposed electrical transport experiments to observe
both the braiding and its non-Abelian statistics.
Considering that the pursuit of topological states
which support quasiparticle excitations obeying non-Ablian statistics
is one of the most important issues in topological quantum computation,
the ability to realize the non-Abelian braiding of chiral Majarona fermions
in our proposed devices is remarkable.
Moreover, for any required unitary transformations, the
braiding operations can be carried out by directly
extending the hybrid TSC/QAHI devices and
can well be controlled and tuned by the electrical method.
This braiding scheme provide a convincing signature of chiral Majarona fermions
and pave a feasible way towards the topological quantum computation.

\noindent\\
\textbf{Acknowledgments}\\
This work was financially supported by National Key R and D Program of China (2017YFA0303301),
NBRP of China (2015CB921102), NSF-China (Grants No. 11574007), and
the Key Research Program of the Chinese Academy of Sciences (Grant No. XDPB08-4).

\clearpage
\appendix
\setcounter{equation}{0}
\setcounter{figure}{0}
\renewcommand{\figurename}{FIG. S\arabic{figure}}
\renewcommand\thefigure{}
\textbf{Supplementary Materials }\\
\section{\uppercase\expandafter{\romannumeral1}. Scattering matrix for a system consisting of two chiral Majorana fermions coupled to a Majorana zero mode}

Here we derive the scattering matrix of the chiral Majorana fermions (CMFs) in Fig.S1(a) (repeated from
the main text for clarity) arising from the coupling with a Majorana zero mode (MZM). The properties of the
CMFs are determined by the Hamiltonian $H_1$ defined in Eq.(2) of the main text.
\begin{figure}[!ht]
\centering
\includegraphics[width=0.5\textwidth]{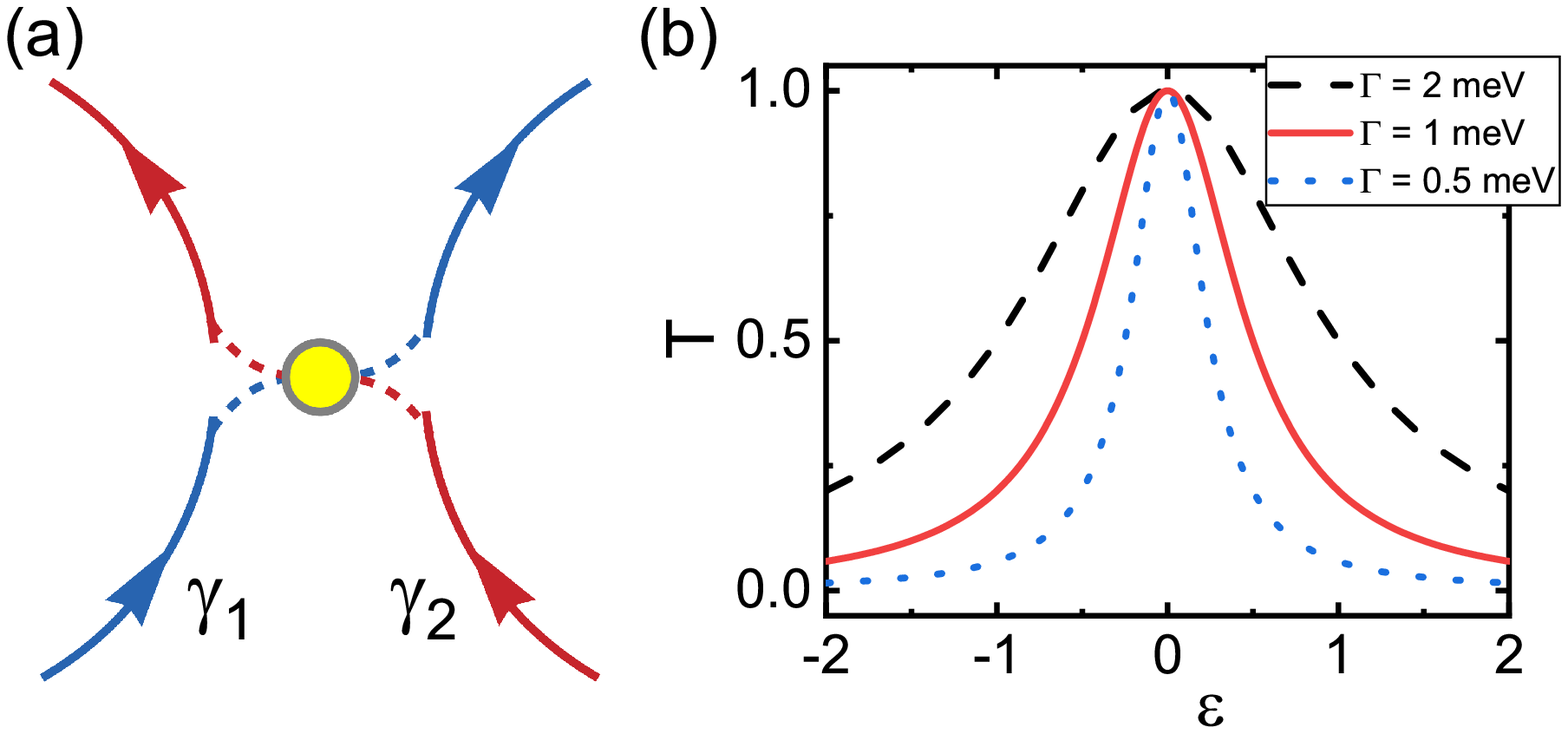}
\caption{(a) Two CMFs $\gamma_1$ and $\gamma_2$ (blue and red arrows) are coupled to a MZM (yellow solid circle).
(b) Transmission coefficient $T(\varepsilon) $ for the scattering between $\gamma_1$ and $\gamma_2$ for different coupling
strength $\Gamma_1=\Gamma_2=\Gamma$.
}
\end{figure}
We begin by investigating the equations of motion for the field operators of the Majorana fermions, i.e.,
$i\partial_t\gamma_{\mathrm{i}}(t)=[\gamma_\mathrm{i}(t),H_1]$ ($\mathrm{i}=0,1,2$) in the Heisenberg picture.
Then the operators obey the differential equations of motion
\begin{eqnarray}
  i\partial_t\gamma_\alpha(x,t) &=& 2i\nu\partial_x\gamma_\alpha(x,t)+it_\alpha\delta(x)\gamma_0(t),\label{s1} \\
  i\partial_t\gamma_0(t) &=& -\sum_{\alpha=1,2}it_\alpha\gamma_\alpha(0,t)\label{s2}.
\end{eqnarray}
Note that we have used the anticommutation relations $\{\gamma_\alpha(x),\gamma_\beta(x')\}=\delta_{\alpha\beta}\delta_{xx'}$ and $\{\gamma_0,\gamma_0\}=1$ to derive these equations.
By the Fourier transformation of $\gamma_\alpha(x,t)$ and $\partial_t\gamma_\alpha(x,t)$
\begin{eqnarray}
  \gamma_\alpha(x,\varepsilon) &=& \int_{-\infty}^{+\infty}\gamma_\alpha(x,t)e^{i\varepsilon t}dt,\label{F1} \\
  \varepsilon\gamma_\alpha(x,\varepsilon) &=& \int_{-\infty}^{+\infty}
   \left[i\partial_t\gamma_\alpha(x,t)\right] e^{i\varepsilon t} dt\label{F2},
\end{eqnarray}
the differential equations (\ref{s1}) and (\ref{s2}) become
\begin{eqnarray}
  \varepsilon\gamma_\alpha(x,\varepsilon) &=& 2i\nu\partial_x\gamma_\alpha(x,\varepsilon)+it_\alpha\delta(x)\gamma_0(\varepsilon),\label{s3} \\
  \varepsilon\gamma_0(\varepsilon) &=& -\sum_{\alpha=1,2}it_\alpha\gamma_\alpha(0,\varepsilon)\label{s4}.
\end{eqnarray}
Integrating the both side of Eq.(\ref{s3}) from $0^-$ to $0^+$, we obtain
\begin{equation}\label{s5}
  2i\nu[\gamma_\alpha(0^+)-\gamma_\alpha(0^-)]+it_\alpha\gamma_0=0.
\end{equation}
Here, the variable $\varepsilon$ has been left out for simplicity.
Using Eq.(\ref{s4}) and $\gamma_\alpha(0)=\frac{\gamma_\alpha(0^+)+\gamma_\alpha(0^-)}{2}$,
one arrives at
\begin{widetext}
\begin{eqnarray}
  (2i\nu+\frac{t_1^2}{2\varepsilon})\gamma_1(0^+)+\frac{t_1t_2}{2\varepsilon}\gamma_2(0^+) &=& (2i\nu-\frac{t_1^2}{2\varepsilon})\gamma_1(0^-)-\frac{t_1t_2}{2\varepsilon}\gamma_2(0^-),\label{s6} \\
  \frac{t_1t_2}{2\varepsilon}\gamma_1(0^+)+(2i\nu+\frac{t_2^2}{2\varepsilon})\gamma_2(0^+) &=& -\frac{t_1t_2}{2\varepsilon}\gamma_1(0^-)+(2i\nu-\frac{t_2^2}{2\varepsilon})\gamma_2(0^-).\label{s7}
\end{eqnarray}
\end{widetext}
Denoting the incoming and outgoing
scattering states of the CMFs by $\gamma_{1/2}(0^-)$ and $\gamma_{1/2}(0^+)$,
respectively, the scattering matrix can be written as
\begin{equation}\label{SM1}
  \left(
    \begin{array}{c}
      \gamma_1(0^+) \\
      \gamma_2(0^+) \\
    \end{array}
  \right)=S_\mathrm{M}\left(
               \begin{array}{c}
                 \gamma_1(0^-) \\
                 \gamma_2(0^-) \\
               \end{array}
             \right).
\end{equation}
Solving from Eqs.(\ref{s6} and \ref{s7}), we get the scattering matrix $S_\mathrm{M}$ as shown in Eq.(4)
in the main text:
\begin{equation}\label{SM2}
  S_\mathrm{M}=\frac{1}{A}\left(
        \begin{array}{cc}
          4i\varepsilon\nu+t_2^2-t_1^2 & -2t_1t_2 \\
          -2t_1t_2 & 4i\varepsilon\nu+t_1^2-t_2^2 \\
        \end{array}
      \right),
\end{equation}
in which $A=4i\varepsilon\nu+t_1^2+t_2^2$ and $\varepsilon$ is the incident energy.
The off-diagonal elements of $S_\mathrm{M}$ matrix correspond to the amplitude for transmission between
$\gamma_1$ and $\gamma_2$. The resulting transmission coefficient is
$T(\varepsilon)= |S_{\mathrm{M},12}|^2 =\frac{4\Gamma_1\Gamma_2}{4\varepsilon^2+(\Gamma_1+\Gamma_2)^2}$
with $\Gamma_\alpha \equiv t_\alpha^2/(2\nu)$
which is same with the Breit-Winger formula describing the resonant scattering
of the lead/QD/lead system.
For large $\varepsilon$,
the transmission between $\gamma_1$ and $\gamma_2$ is forbidden ($T=0$),
while as $\varepsilon$ approaching zero,
a resonant exchange occurs which transform $\gamma_1\rightarrow-\gamma_2$
and $\gamma_2\rightarrow-\gamma_1$ ($T=1$) as shown in Fig.S1(b).

\section{\uppercase\expandafter{\romannumeral2}. Scattering matrix for a system consisting of two chiral Majorana fermions coupled to a quantum dot}

\begin{figure}[!ht]
\centering
\includegraphics[width=0.2\textwidth]{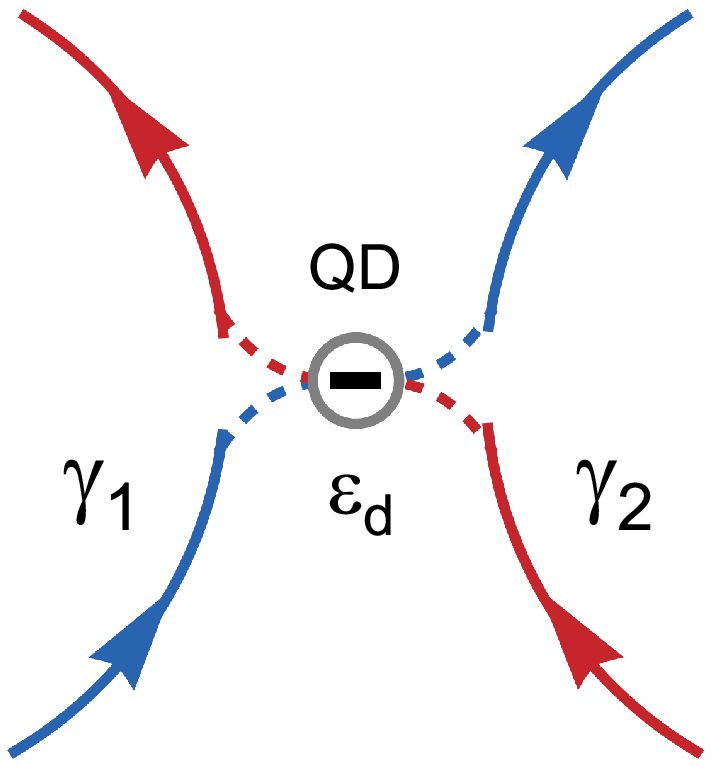}
\caption{Two CMFs $\gamma_1$ and $\gamma_2$ (blue and red arrows) are coupled to a QD.
}
\end{figure}

Bellow, we derive the scattering matrix of the CMFs in Fig.S2 (repeated from
the main text for clarity) arising from the coupling with a quantum dot (QD).
In this situation, the equation of motion for the field operators is $i\partial_tO(t)=[O(t),H_2]$, where $O(t)$ stands for the field operators of CMFs and the
fermion state inside the QD in Heisenberg picture, and $H_2$ is defined in Eq.(5) of the main text. Here, the differential equations for the operators become
\begin{eqnarray*}
  i\partial_t\gamma_\alpha(x,t) &=& 2i\nu\partial_x\gamma_\alpha(x,t)+i\frac{\tilde{t}_\alpha}{\sqrt{2}}\delta(x)[d(t)+d^\dag(t)],\label{s8} \\
  i\partial_td(t) &=& \varepsilon_dd -\sum_{\alpha=1,2}i\frac{\tilde{t}_\alpha}{2}\gamma_\alpha(0,t),\label{s9}\\
  i\partial_td^\dag(t) &=& -\varepsilon_dd^\dag -\sum_{\alpha=1,2}i\frac{\tilde{t}_\alpha}{\sqrt{2}}\gamma_\alpha(0,t)\label{s10}.
\end{eqnarray*}
After a Fourier transformation similar to Eq.(\ref{F1}) and (\ref{F2}), we arrive at
\begin{eqnarray}
  \varepsilon\gamma_\alpha(x,\varepsilon) &=& 2i\nu\partial_x\gamma_\alpha(x,\varepsilon)+i\frac{\tilde{t}_\alpha}{\sqrt{2}}\delta(x)\gamma_0(\varepsilon),\label{s11} \\
  \varepsilon d(\varepsilon) &=& \varepsilon_dd(\varepsilon)-\sum_{\alpha=1,2}i\frac{\tilde{t}_\alpha}{\sqrt{2}}\gamma_\alpha(0,\varepsilon)\label{s12},\\
  \varepsilon d^\dag(\varepsilon) &=& \varepsilon_dd^\dag(\varepsilon)-\sum_{\alpha=1,2}i\frac{\tilde{t}_\alpha}{\sqrt{2}}\gamma_\alpha(0,\varepsilon)\label{s13}.
\end{eqnarray}
Integrating the both side of Eq.(\ref{s11}) from $0^-$ to $0^+$, we obtain
\begin{equation}\label{s14}
  2i\nu[\gamma_\alpha(0^+)-\gamma_\alpha(0^-)]+i\frac{\tilde{t}_\alpha}{\sqrt{2}}(d+d^\dag)=0.
\end{equation}
Following the same procedure in the derivation of Eqs. (\ref{s6},\ref{s7},\ref{SM1}), the scattering matrix $S_\mathrm{Q}$ for the QD coupling case can be found
\begin{equation}\label{SQ}
  S_\mathrm{Q}=\frac{1}{\widetilde{A}}\left(
        \begin{array}{cc}
          \widetilde{B}+\tilde{t}_2^2-\tilde{t}_1^2 & -2\tilde{t}_1\tilde{t}_2 \\
          -2\tilde{t}_1\tilde{t}_2 & \widetilde{B}+\tilde{t}_1^2-\tilde{t}_2^2 \\
        \end{array}
      \right),
\end{equation}
where $\widetilde{B}=4i\nu(\varepsilon^2-\varepsilon_d^2)/\varepsilon$ and  $\widetilde{A}=\widetilde{B}+\tilde{t}_1^2+\tilde{t}_2^2$.
By setting $\varepsilon_d=0$, the scattering matrix $S_\mathrm{Q}$ recovers $S_\mathrm{M}$ in Eq. (\ref{SM2}). For finite $\varepsilon_d$, the transmission coefficient now becomes
$\widetilde{T}(\varepsilon) = |S_{\mathrm{Q},12}|^2 = \frac{4\widetilde{\Gamma}_1\widetilde{\Gamma}_2}
{4(\varepsilon^2-\varepsilon_d^2)^2+(\widetilde{\Gamma}_1+\widetilde{\Gamma}_2)^2}$
with $\widetilde{\Gamma}_\alpha \equiv \tilde{t}_\alpha^2/(2\nu)$.

\section{\uppercase\expandafter{\romannumeral3}. Calculation the transmission coefficients and terminal conductance in proposed devices}

In this supplementary section, we give a detailed calculation of the transmission coefficients and terminal conductances of the proposed
experimental devices for the observation of the braiding and its non-Abelian statistics.

\subsection{A. Device that allows the observation of the braiding of chiral Majorana
fermions }

\begin{figure}[!ht]
\centering
\includegraphics[width=0.25\textwidth]{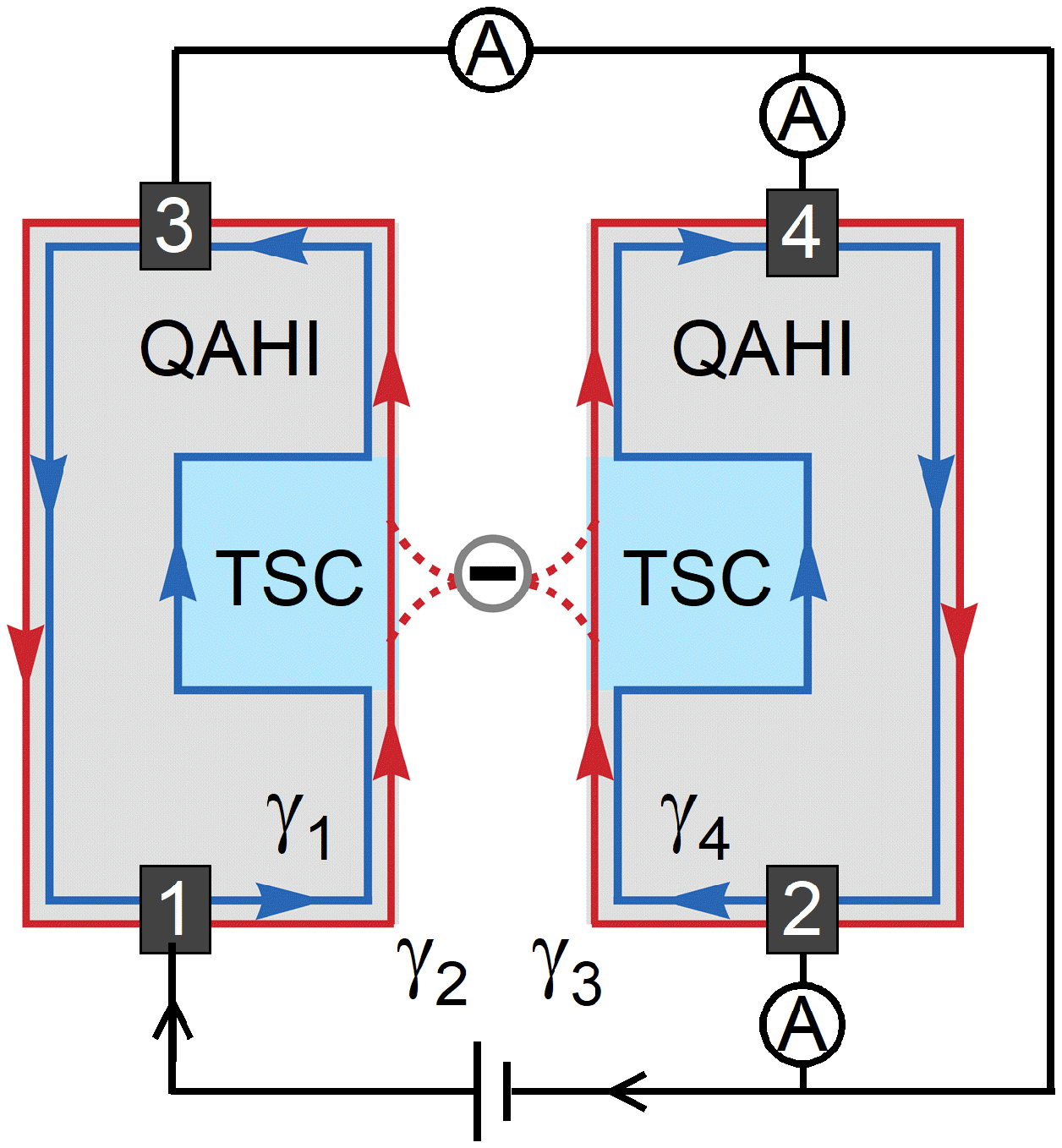}
\caption{Experimental device that allows the observation of the braiding of chiral Majorana fermions based on TSC/QAHI junctions.
Transport process via four chiral Majorana fermions $\gamma_i$. The chirality of $\gamma_i$ in left junction is different with
the one in right junction resulting from the opposite magnetization in two regions. The QD coupled to the TSC/QAHI junctions can trigger
the exchange of $\gamma_2$ and $\gamma_3$. The coupling strength is $\widetilde{\Gamma}_1=\widetilde{\Gamma}_2=\Gamma$.
}
\end{figure}

First, we consider the device that allows the observation of the braiding of CMFs based on TSC/QAHI junctions in Fig.S3 (repeated from
the main text for clarity). In the transport process of the four CMFs $\gamma_i$ ($i=1,2,3,4$) in Fig.S3, $\gamma_1$ and $\gamma_4$
are transported directly into lead 3 and lead 4 while $\gamma_2$ and $\gamma_3$ are scattered by the QD. For simplicity, denoting the incoming CMFs by $\gamma_2$ and $\gamma_3$, and outgoing scattering states by $\gamma_2'$ and $\gamma_3'$ respectively, the scattering matrix  $S_\mathrm{Q}$ can be written as
\begin{equation}\label{SQ3}
  \left(
    \begin{array}{c}
      \gamma_2' \\
      \gamma_3' \\
    \end{array}
  \right)=S_\mathrm{Q}\left(
               \begin{array}{c}
                 \gamma_2 \\
                 \gamma_3 \\
               \end{array}
             \right),
\end{equation}
where $S_\mathrm{Q}$ is given by Eq. (\ref{SQ}).
We define four operators as
\begin{eqnarray}
  a_1 &=& (\gamma_1+i\gamma_2)/\sqrt{2}, \\
  a_2 &=& (\gamma_3+i\gamma_4)/\sqrt{2}, \\
  b_3 &=& (\gamma_1+i\gamma_2')/\sqrt{2}, \\
  b_4 &=& (\gamma_3'+i\gamma_4)/\sqrt{2},
\end{eqnarray}
which represent the incoming QAHI edge modes ($a_1$ from lead 1 and $a_2$ from lead 2) and outgoing QAHI edge modes ($b_3$ toward to lead 3 and $b_4$ toward to lead 4), respectively. By using Eq. (\ref{SQ3}), we can obtain
\begin{widetext}
\begin{eqnarray}
  b_3 &=& [(1+S_{\mathrm{Q},11})a_1+(1-S_{\mathrm{Q},11})a_1^\dag+iS_{\mathrm{Q},12}a_2+iS_{\mathrm{Q},12}a_2^\dag]/2, \\
  b_3^\dag &=& [(1-S_{\mathrm{Q},11}^*)a_1+(1+S_{\mathrm{Q},11}^*)a_1^\dag-iS_{\mathrm{Q},12}^*a_2-iS_{\mathrm{Q},12}^*a_2^\dag]/2, \\
  b_4 &=& [-iS_{\mathrm{Q},21}a_1+iS_{\mathrm{Q},21}a_1^\dag+(1+S_{\mathrm{Q},22})a_2+(-1+S_{\mathrm{Q},22})a_2^\dag]/2, \\
  b_4^\dag &=& [-iS_{\mathrm{Q},21}^*a_1+iS_{\mathrm{Q},21}^*a_1^\dag+(-1+S_{\mathrm{Q},22}^*)a_2+(1+S_{\mathrm{Q},22}^*)a_2^\dag]/2.
\end{eqnarray}
\end{widetext}
These equations can be rewritten into a compact form as
\begin{equation}
  \left(
     \begin{array}{c}
       b_3 \\
       b_3^\dag \\
       b_4 \\
       b_4^\dag\\
     \end{array}
   \right)=\overrightarrow{S}\left(
                                          \begin{array}{c}
                                            a_1 \\
                                            a_1^\dag \\
                                            a_2 \\
                                            a_2^\dag \\
                                          \end{array}
                                        \right),
\end{equation}
with
\begin{equation}
  \overrightarrow{S}=\frac{1}{2}\left(
                       \begin{array}{cccc}
                         1+S_{\mathrm{Q},11} & 1-S_{\mathrm{Q},11} & iS_{\mathrm{Q},12} & iS_{\mathrm{Q},12} \\
                         1-S_{\mathrm{Q},11}^* & 1+S_{\mathrm{Q},11}^* & -iS_{\mathrm{Q},12}^* & -iS_{\mathrm{Q},12}^* \\
                         -iS_{\mathrm{Q},21} & iS_{\mathrm{Q},21} & 1+S_{\mathrm{Q},22} & -1+S_{\mathrm{Q},22} \\
                         -iS_{\mathrm{Q},21}^* & iS_{\mathrm{Q},21}^* & -1+S_{\mathrm{Q},22}^* & 1+S_{\mathrm{Q},22}^* \\
                       \end{array}
                     \right).
\end{equation}
Then the transport coefficients for normal tunneling and Andreev reflection from lead 1 to lead 3 and lead 4 can
be obtained by
\begin{eqnarray}
  T_{31}(\varepsilon) &=& |\overrightarrow{S}_{11}|^2 = |1+S_{\mathrm{Q},11}|^2/4, \label{T1}\\
  T_{31}^A(\varepsilon) &=& |\overrightarrow{S}_{21}|^2 = |1-S_{\mathrm{Q},11}^*|^2/4,\label{T2} \\
  T_{41}(\varepsilon) &=& |\overrightarrow{S}_{31}|^2 = |iS_{\mathrm{Q},21}|^2/4, \label{T3}\\
  T_{41}^A(\varepsilon) &=& |\overrightarrow{S}_{41}|^2 = |iS_{\mathrm{Q},21}^*|^2/4\label{T4}.
\end{eqnarray}
Considering the finite temperature $\mathcal{T}$,
the effective transmission coefficients can be derived
\begin{eqnarray}
  T_{31} &=& \int_{-\infty}^{+\infty}T_{31}(\varepsilon)(-\frac{\partial f}{\partial\varepsilon})d\varepsilon,\label{T5} \\
  T_{31}^A &=& \int_{-\infty}^{+\infty}T_{31}^A(\varepsilon)(-\frac{\partial f}{\partial\varepsilon})d\varepsilon,\label{T6} \\
  T_{41} &=& \int_{-\infty}^{+\infty}T_{41}(\varepsilon)(-\frac{\partial f}{\partial\varepsilon})d\varepsilon,\label{T7} \\
  T_{41}^A &=& \int_{-\infty}^{+\infty}T_{41}^A(\varepsilon)(-\frac{\partial f}{\partial\varepsilon})d\varepsilon\label{T8},
\end{eqnarray}
in which $f(\varepsilon)=[\mathrm{exp}(\varepsilon/k_B\mathcal{T}+1]^{-1}$ is the Fermi distribution
function.

Next, we assume that the voltage of lead 1 is set to be $V$
and the voltage of lead 2,3,4 are the same assumed to be $U$.
The measured current in lead $n$ can be calculated using
the multiprobe Landauer-B\"{u}ttiker formula\cite{Sun1,Sun2,datta}
\begin{equation}\label{Cur}
I_n=\frac{e^2}{h}\sum_mT_{nm}(V_n-V_m)+T_{nm}^A(V_n+V_m),
\end{equation}
where $T_{nm}$ ($T_{nm}^A$) is the normal tunneling (Andreev reflection) coefficient from lead $m$
to lead $n$. The nonvanishing coefficients are
$T_{13}=T_{24}=1$ and $T_{nm}$ ($T_{nm}^A$) with $n=3,4$ and $m=1,2$ which can be calculated from the scattering matrix $\overrightarrow{S}$
as in Eqs.(\ref{T5}-\ref{T8}). With these coefficients, the currents of the leads are
\begin{eqnarray}
  I_1 &=& \frac{e^2}{h}T_{13}(V-U), \label{II1} \\
  I_2 &=& 0, \label{II2}\\
  I_3 &=& \frac{e^2}{h}[T_{31}(U-V)+T_{31}^A(U+V)+2T_{32}^AU], \label{II3}\\
  I_4 &=& \frac{e^2}{h}[T_{41}(U-V)+T_{41}^A(U+V)+2T_{42}^AU]. \label{II4}
\end{eqnarray}
Then by the current conservation $I_1+I_2+I_3+I_4=0$,
one gets
\begin{eqnarray}
U =\frac{T_{31} -T_{31}^A  +T_{41} -T_{41}^A -T_{13}}
 {T_{31} +T_{31}^A +2T_{32}^A + T_{41} +T_{41}^A +2T_{42}^A  - T_{13}} V. \label{UV1}
\end{eqnarray}
The conductance of lead $n$ is defined as
$G_n=\frac{-I_n}{V_1-V_n} =\frac{-I_n}{V-U}$, which can be obtained from Eqs.(\ref{II1}-\ref{UV1})
straightforwardly.

\subsection{B. Device for the observation of non-Abelian statistics}
We now calculate the transport properties of the four CMFs after sequential exchanges. If the final state depends on the braiding
order as well as the measurement of terminal conductance, this means that the braiding of CMFs obeys the non-Abelian statistics.
First, we consider the joint operations with two exchanges, $\sigma_2\sigma_1$ and $\sigma_1\sigma_2$. $\sigma_2\sigma_1$ will transform the
four CMFs according to ($\gamma_1,\gamma_2,\gamma_3,\gamma_4$)$\rightarrow$($-\gamma_2,-\gamma_3,\gamma_1,\gamma_4$). After the braiding, the outgoing
QAHI edge modes ($b_3$ toward to lead 3 and $b_4$ toward to lead 4) now become
\begin{eqnarray*}
  b_3 &=& (-\gamma_2-i\gamma_3)/\sqrt{2}=(ia_1-ia_1^\dag-ia_2-ia_2^\dag)/2, \\
  b_3^\dag &=& (-\gamma_2+i\gamma_3)/\sqrt{2}=(ia_1-ia_1^\dag+ia_2+ia_2^\dag)/2, \\
  b_4 &=& (\gamma_1+i\gamma_4)/\sqrt{2}=(a_1+a_1^\dag+a_2-a_2^\dag)/2, \\
  b_4^\dag &=& (\gamma_1-i\gamma_4)/\sqrt{2}=(a_1+a_1^\dag-a_2+a_2^\dag)/2,
\end{eqnarray*}
in which we have used $\gamma_1=\frac{1}{\sqrt{2}}(a_1+a_1^\dag),\gamma_2=\frac{1}{i\sqrt{2}}(a_1-a_1^\dag),\gamma_3=\frac{1}{\sqrt{2}}(a_2+a_2^\dag),\gamma_4=\frac{1}{i\sqrt{2}}(a_2-a_2^\dag)$.
Now, the scattering matrix $\overrightarrow{S}$ relating the incoming modes $a_{1,2},a_{1,2}^\dag$ and $b_{3,4},b_{3,4}^\dag$ becomes
\begin{equation*}
 \overrightarrow{S}=\frac{1}{2}\left(
                      \begin{array}{cccc}
                        i & -i & -i & -i \\
                        i & -i & i & i \\
                        1 & 1 & 1 & -1 \\
                        1 & 1 & -1 & 1 \\
                      \end{array}
                    \right),
\end{equation*}
which gives $T_{31}=T_{41}=T_{31}^A=T_{41}^A=1/4$ and $T_{32}=T_{42}=T_{32}^A=T_{42}^A=1/4$.
By using these normal tunneling and Andreev reflection coefficients,
the conductances can be obtained from Eqs.(\ref{II1}-\ref{UV1}),
and the results are that $G_3=G_4=\frac{e^2}{2h}$.
Similarly, the scattering matrix for the joint operation $\sigma_1\sigma_2$
can be obtained following the same procedure as
\begin{equation*}
 \overrightarrow{S}=\frac{1}{2}\left(
                      \begin{array}{cccc}
                        i & i & 1 & 1 \\
                        -i & -i & 1 & 1 \\
                        -i & i & 1 & -1 \\
                        -i & i & -1 & 1 \\
                      \end{array}
                    \right),
\end{equation*}
thus it gives the same terminal conductance $G_3=G_4=\frac{e^2}{2h}$ as $\sigma_2\sigma_1$ despite that the final states are different.

Next, we turn to the joint operators $\sigma_2\sigma_1\sigma_2$ and $\sigma_2\sigma_2\sigma_1$. After the transformation by $\sigma_2\sigma_1\sigma_2$,
($\gamma_1,\gamma_2,\gamma_3,\gamma_4$) becomes ($\gamma_3,-\gamma_2,\gamma_1,\gamma_4$), and the resulting scattering matrix is
\begin{equation*}
 \overrightarrow{S}=\frac{1}{2}\left(
                      \begin{array}{cccc}
                        -1 & 1 & 1 & 1 \\
                        1 & -1 & 1 & 1 \\
                        1 & 1 & 1 & -1 \\
                        1 & 1 & -1 & 1 \\
                      \end{array}
                    \right).
\end{equation*}
We obtain $T_{31}=T_{41}=T_{31}^A=T_{41}^A=1/4$ and $T_{32}=T_{42}=T_{32}^A=T_{42}^A=1/4$, and $G_3=G_4=\frac{e^2}{2h}$ again.
However, the joint operator $\sigma_2\sigma_2\sigma_1$ transforms the CMFs $\gamma_i$ according to
($\gamma_1,\gamma_2,\gamma_3,\gamma_4$)$\rightarrow$($-\gamma_2,-\gamma_1,-\gamma_3,\gamma_4$). The related scattering matrix now becomes
\begin{equation*}
 \overrightarrow{S}=\frac{1}{2}\left(
                      \begin{array}{cccc}
                        0 & -i & 0 & 0 \\
                        i & 0 & 0 & 0 \\
                        0 & 0 & 0 & -1 \\
                        0 & 0 & 1 & 0 \\
                      \end{array}
                    \right),
\end{equation*}
which gives $T_{31}^A=T_{42}^A=1$ and $T_{31}=T_{41}=T_{41}^A=T_{32}=T_{42}=T_{42}^A=0$. The corresponding terminal conductances are $G_4=\frac{e^2}{h}$
and $G_3=0$. At this point, we can see that not only the two joint operators $\sigma_2\sigma_1\sigma_2$ and $\sigma_2\sigma_2\sigma_1$ drive the CMFs $\gamma_i$
from the same initial state into different final states, but also the resulting terminal conductance is distinct. These results provide a method to observe the non-Abelian
braiding statistics of the CMFs.

\section{\uppercase\expandafter{\romannumeral4}. An universal device for executing various braiding operations in an electrically controllable manner}

\begin{figure}
\centering
\includegraphics[width=0.3\textwidth]{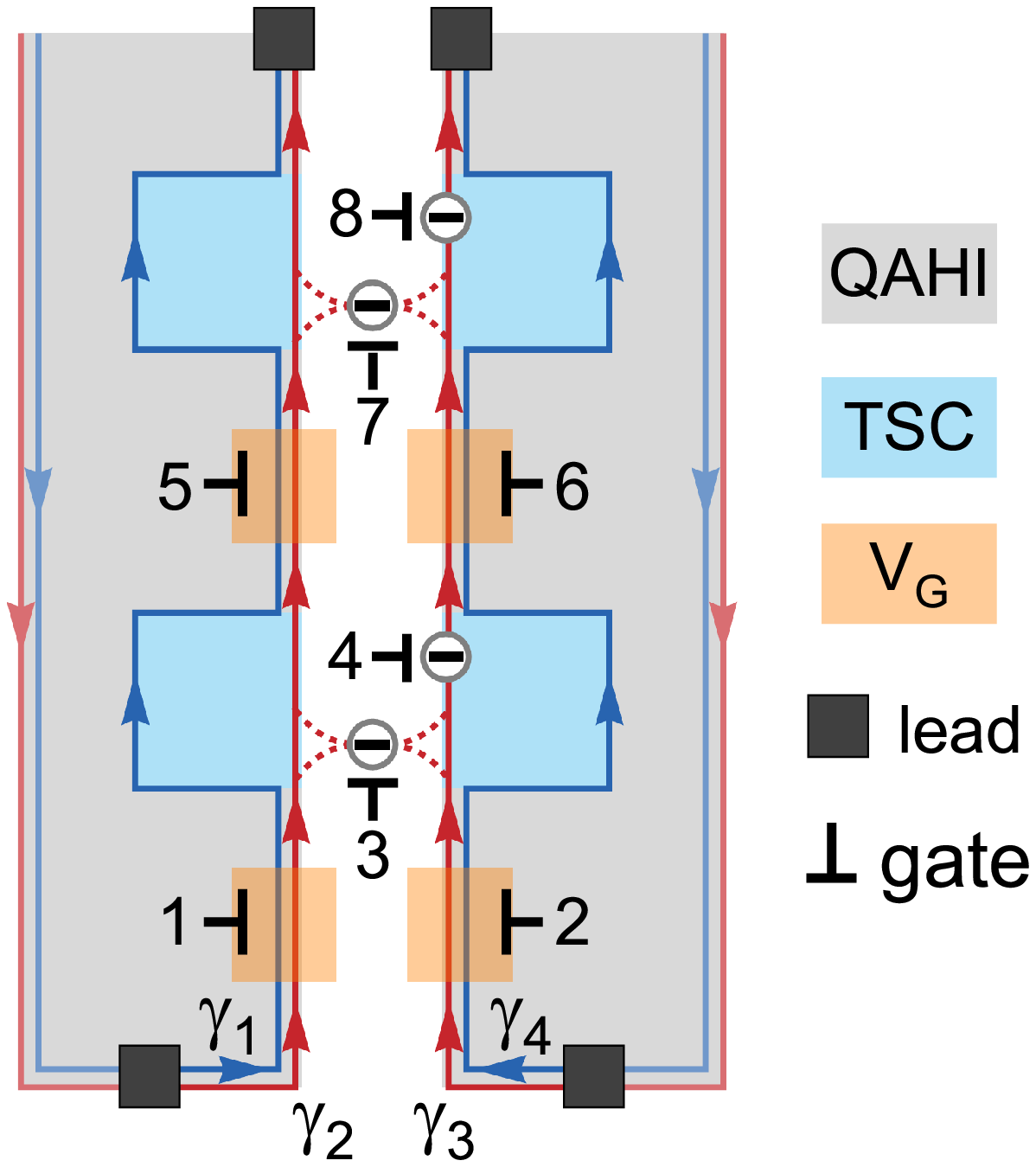}
\caption{An universal device that allows various braiding operations by tuning the gates based on TSC/QAHI junctions.
The pair of QDs between the two TSC regions are used to carry out $\sigma_2$ as discussed in the main text.
The gate voltages $V_G$ placed on the QAHI edge are used to carry out $\sigma_1$ and $\sigma_3$. The QDs and gated reigons
are controlled by gates.
}
\end{figure}

As discussed in the main text, the coupling with QDs provides an electrical control of the braiding of CMFs.
In this final section, we further propose an universal device for executing various braiding operations
controlled by gates as shown in Fig.S4 to show the advantages of our strategy. Similar to the devices in Fig.4
of the main text, the device as shown in Fig.S4 is made by two ribbons of hybrid TSC/QAHI junctions with opposite
out-plane magnetization. Here, we introduce four QAHI regions covered by gate voltage $V_G$ as denoted by orange
shadow regions to carry out the braiding operator $\sigma_1$ ($\sigma_3$) controlled by gate 1 and 5 (2 and 6)\cite{LianB}.
Moreover, four QDs controlled by gate 3,4,7,8 are placed between the TSC regions to carry out the braiding operator $\sigma_2$.
If all the operators controlled by the gates are working, the operation realized in Fig.S4
is $\sigma_2\sigma_3\sigma_1\sigma_2\sigma_3\sigma_1$ and transforms the four CMFs ($\gamma_1,\gamma_2,\gamma_3,\gamma_4$)
into ($-\gamma_4,\gamma_3,-\gamma_2,\gamma_1$). The advantage of our proposed device is that all the modules to
realize the braiding operators are electrically controllable and this point makes various operations available
in a single device by tuning the gates.

\begin{figure}[!ht]
\centering
\includegraphics[width=0.48\textwidth]{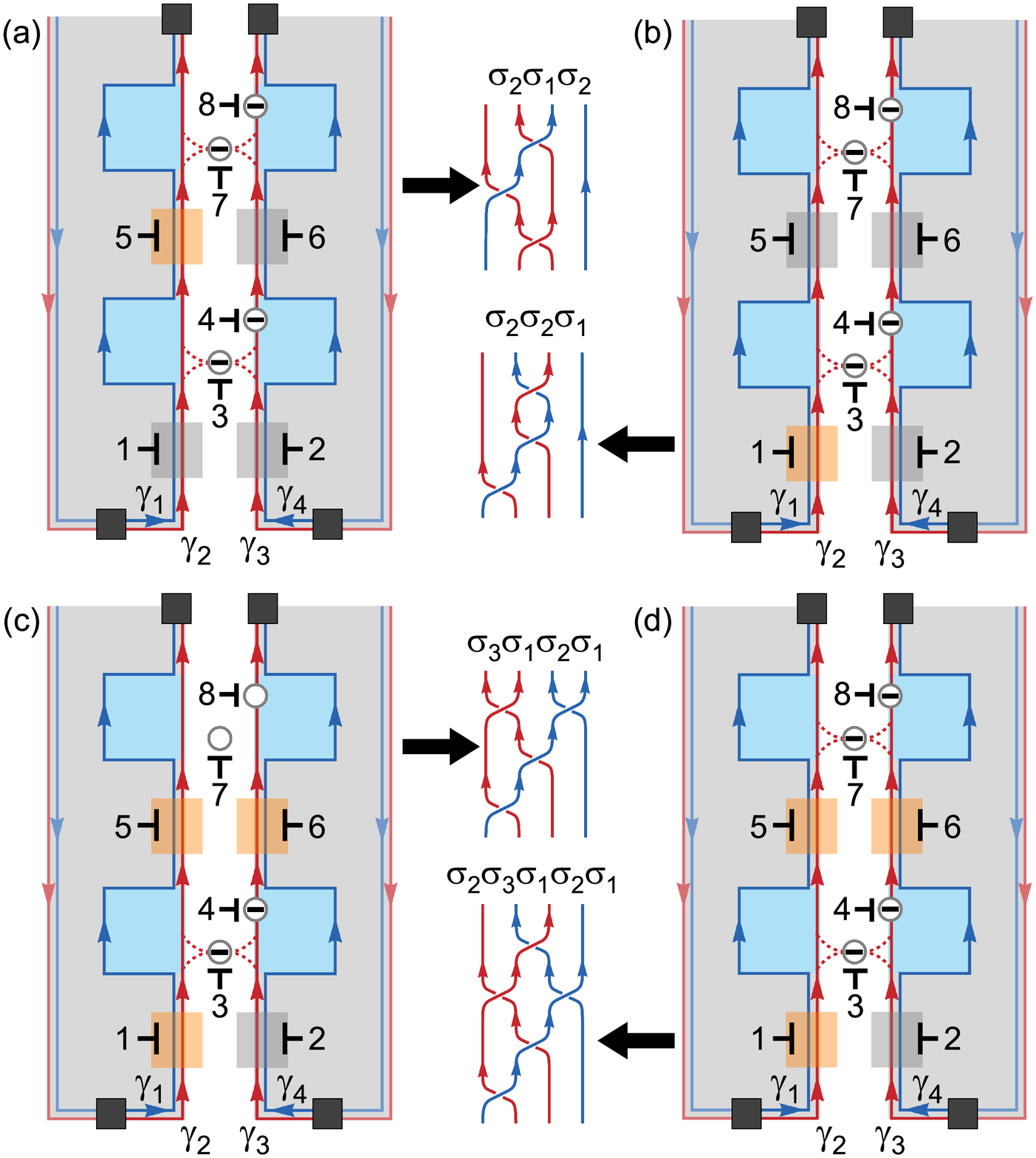}
\caption{Four representative operations realized in the device from Fig.S4 by tuning the gates. (a) By tuning the gates
1,2,6 away from the working conditions as represented by dark gray shadow regions, the resulting operation recovers $\sigma_2\sigma_1\sigma_2$ in Fig.4(a) of the main text. (b) the operation $\sigma_2\sigma_2\sigma_1$ by tuning gates 2,5,6,
reminiscent of Fig.4(b) in the main text. (c) and (d) the operations $\sigma_3\sigma_1\sigma_2\sigma_1$ and $\sigma_2\sigma_3\sigma_1\sigma_2\sigma_1$. The hollow circles in (c) means that the energy level $\varepsilon_d$ is tuned away
from zero by gate 7,8 and the QDs cannot braid the CMFs anymore. The middle part shows the corresponding braiding of $\gamma_i$ ($i= 1, 2, 3, 4$).
}
\end{figure}

In principle, there are 64 ($2^6$) combinations of $\sigma_i$ ($i=1,2,3$) which can be executed by the device in Fig.S4.
Here, we show four representative operations in Fig.S5. If one tunes the gates 1,2,6 away from the working conditions
for the braiding operators $\sigma_1$ and $\sigma_3$, then the resulting operation is $\sigma_2\sigma_1\sigma_2$ as
shown in Fig.S5(a), i.e., the joint operation in Fig.4a of the main text. Similarly, the joint operation
$\sigma_2\sigma_2\sigma_1$ in Fig.4b of the main text can be carried out by tuning the gates 2,5,6 away from the
working conditions as shown in Fig.S5(b). Moreover, as shown in the main text, if the energy level $\varepsilon_d$
of the QDs deviates from zero, the braiding of CMFs is unrealizable as indicated by hollow circle in Fig.S5(c).
In Fig.S5(c), we disable the QDs controlled by gate 7,8 and the $V_G$ controlled by gate 2, then the consequent
operation $\sigma_3\sigma_1\sigma_2\sigma_1$ transform ($\gamma_1,\gamma_2,\gamma_3,\gamma_4$)
into ($\gamma_3,-\gamma_2,-\gamma_4,\gamma_1$). If we switch on the QD again as shown in Fig.S5(d), another operation $\sigma_2\sigma_3\sigma_1\sigma_2\sigma_1$
is achieved and the CMFs ($\gamma_1,\gamma_2,\gamma_3,\gamma_4$) are transformed to ($\gamma_3,\gamma_4,-\gamma_2,\gamma_1$).
The remaining operations can be also achieved by tuning the corresponding gates.
\end{document}